\def\papertitle{Neural Morphing: Sequence-Optimized \\ Token-Level Morphing in Neural Audio Codecs}
\def\paperauthorA{Emmanouil Karystinaios}
\def\paperauthorB{Author Two}
\def\paperauthorC{Author Three}
\def\paperauthorD{Author Four}

\documentclass[twoside,a4paper]{article}
\usepackage{etoolbox}

\usepackage{dafx26demo}

\usepackage{amsmath,amssymb,amsfonts,amsthm}
\usepackage{siunitx}
\usepackage{euscript}
\usepackage[T1]{fontenc}
\usepackage[utf8]{inputenc}
\usepackage{ifpdf}
\usepackage[english]{babel}
\usepackage{caption}
\usepackage{subfig} 
\usepackage{color}
\usepackage{booktabs}
\usepackage{lipsum}
\usepackage{microtype}

\input glyphtounicode
\ninept

\newcounter{numauth}
\newcounter{listcnt}
\newcommand\authcnt[1]{\ifdefined#1 \stepcounter{numauth} \fi}

\newcommand\addauth[1]{
\ifdefined#1 
\stepcounter{listcnt}
\ifnum \value{listcnt}<\value{numauth}
\appto\authorslist{, #1}
\else
\appto\authorslist{~and~#1}
\fi
\fi}
\authcnt{\paperauthorB}
\authcnt{\paperauthorC}
\authcnt{\paperauthorD}
\authcnt{\paperauthorE}
\authcnt{\paperauthorF}
\authcnt{\paperauthorG}
\authcnt{\paperauthorH}
\authcnt{\paperauthorI}
\authcnt{\paperauthorJ}
\def\authorslist{\paperauthorA}
\addauth{\paperauthorB}
\addauth{\paperauthorC}
\addauth{\paperauthorD}
\addauth{\paperauthorE}
\addauth{\paperauthorF}
\addauth{\paperauthorG}
\addauth{\paperauthorH}
\addauth{\paperauthorI}
\addauth{\paperauthorJ}

\usepackage{times}

\newif\ifpdf
\ifx\pdfoutput\relax
\else
   \ifcase\pdfoutput
      \pdffalse
   \else
      \pdftrue
   \fi
\fi

\ifpdf 
  \usepackage[pdftex,
    pdftitle={\papertitle},
    pdfauthor={\authorslist},
    pdfsubject={Proceedings of the 29th International Conference on Digital Audio Effects (DAFx26)},
    colorlinks=false, 
    bookmarksnumbered, 
    pdfstartview=XYZ 
  ]{hyperref}
  \usepackage[pdftex]{graphicx}
\else 
  \usepackage[dvips]{epsfig,graphicx}
  \usepackage[dvips,
    pdftitle={\papertitle},
    pdfauthor={\authorslist},
    pdfsubject={Proceedings of the 29th International Conference on Digital Audio Effects (DAFx26)},
    colorlinks=false, 
    bookmarksnumbered, 
    pdfstartview=XYZ 
  ]{hyperref}
\fi
\usepackage[hypcap=true]{caption}
\title{\papertitle}

\affiliation
{\paperauthorA}
{Institute of Computational Perception \\ Johannes Kepler University Linz, Austria\\
{\tt \href{mailto:emmanouil.karystinaios@jku.at}{emmanouil.karystinaios@jku.at}}
}

\begin{document}
\ifpdf 
  \DeclareGraphicsExtensions{.png,.jpg,.pdf}
\else  
  \DeclareGraphicsExtensions{.eps}
\fi


\maketitle

\begin{abstract}
Neural audio codecs were originally developed for high-fidelity compression; however, their latent token representations and expressive decoders also constitute a powerful substrate for controllable audio transformation. This work introduces Neural Morphing, a training-free token-domain audio effect that selects residual-vector-quantized (RVQ) token grains from a user palette and decodes the edited stream through a pretrained codec. The method combines an RVQ-group transfer policy that separates coarse, middle, and fine codebook groups with a continuity-constrained sequence matcher that replaces independent greedy selection with bounded beam search. The intended output is a controlled hybrid: the source preserves rhythmic organization while the palette contributes timbral color and residual detail. We focus on the implementation and realtime behavior of a deployable VST3/AU system, including chunked rendering, palette-size scaling, and backend health checks.
\end{abstract}

\section{Introduction}

Many creative audio effects aim to move a source toward another sound while preserving enough temporal structure for the result to remain playable. Classical audio morphing often interpolates spectral or parametric features, whereas audio mosaicing selects short units from a corpus and reassembles them into a new texture~\cite{Serra:1990:SpectralInterpolation,Schwarz:2006:CorpusBased}. Neural audio codecs offer a third domain: learned discrete token streams that can be edited and decoded without training a new generator~\cite{Kumar:2023:DAC}. Neural Morphing explores this domain as a deployable digital audio effect for source-conditioned palette transformation.

The creative motivation is deliberately practical. A sound designer may want a drum loop to keep its groove while acquiring the color of metal impacts, vinyl debris, vocal consonants, or field-recorded textures; a performer may want to audition a palette as a timbral brush without training a model for every session; and a producer may want morphing behavior that is repeatable, automatable, and compatible with a DAW workflow. In this setting, the goal is not to reconstruct a target waveform. The target material is a palette of possible local colors, and the source supplies timing, gesture, and large-scale organization.

Technically, the method sits between several established families. It is related to cross-synthesis and spectral morphing because it transfers aspects of one sound onto another; it is related to mosaicing because it selects short units from a corpus under continuity constraints; and it is related to latent granular resynthesis because it matches codec-domain grains before neural decoding~\cite{tokui2025latent}. However, it is not a conventional waveform mosaicing system, since overlap-add waveform grains are never assembled directly, and it is not a trained generative transformation model, since no new model is fitted to the palette. We therefore describe the technique as token-domain palette-based morphing with mosaicing-like sequence selection.

The system treats a palette as a collection of codec-token grains. During playback or rendering, the source is encoded, each source grain retrieves compatible palette candidates, and a sequence optimizer selects a path through the palette before token replacement and neural decoding. This gives the performer familiar controls---palette choice, dry/wet balance, grain size, search width, and structure/detail transfer---while hiding the codec-specific token operations.

This paper focuses on three engineering details: (i) the token-domain components, (ii) the VST3/AU implementation and realtime constraints, and (iii) deployment-oriented checks. Portability was tested informally with other codec integrations, including SpectroStream and CoDiCodec-style adapters; however, the central evidence here uses DAC because it exposes a native multi-codebook RVQ stack suitable for band-aware editing.

\begin{figure*}[t]
    \centering
    \includegraphics[width=\textwidth]{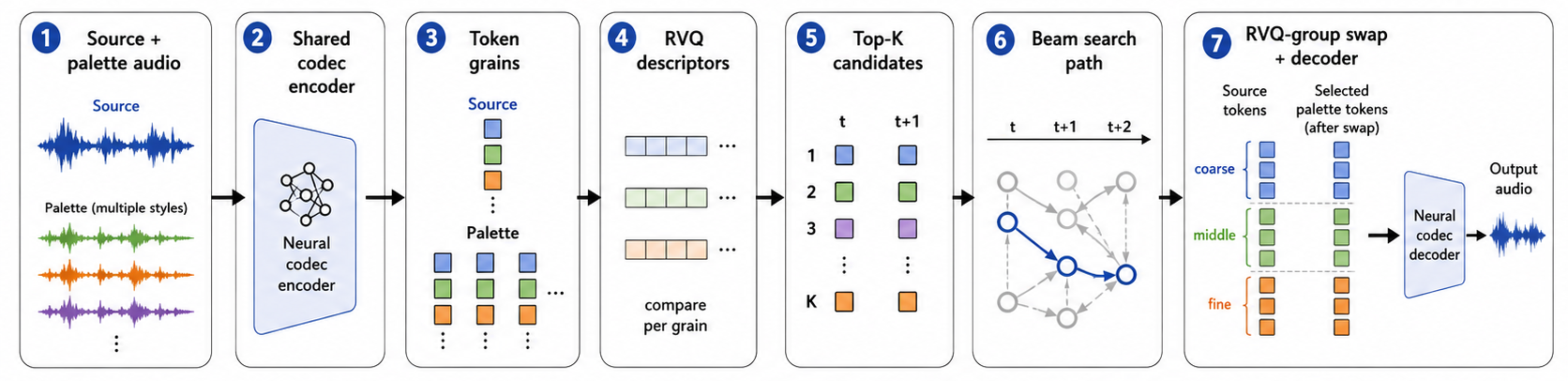}
    \caption{Neural Morphing pipeline. Source and palette audio are encoded into neural-codec tokens, segmented into grains, matched in a codec-induced descriptor space, optimized as a continuity-constrained palette path, edited with full-layer or RVQ-group token transfer, and decoded back to audio.}
    \label{fig:pipeline}
\end{figure*}

\section{System Overview}

Figure~\ref{fig:pipeline} summarizes the processing chain. Let $E$ and $D$ be a neural-codec encoder and decoder. An input waveform $x(t)$ is encoded into token frames
\begin{equation}
    \mathbf{Z}=\{\mathbf{z}_n\}_{n=1}^{N},\qquad
    \mathbf{z}_n=(z_{n,1},\ldots,z_{n,L}),
\end{equation}
where $L$ is the number of residual quantizer layers. DAC is used at 44.1~kHz with nine RVQ codebooks and approximately 87 token frames per second~\cite{Kumar:2023:DAC}. Source and palette streams are segmented into fixed-length token grains of $G$ frames with hop $H$; the tuned DAC setting uses $G=7$ and $H=2$.

Each grain is represented by a codec-induced descriptor. When codebook embeddings are available, selected embeddings are averaged over the grain and concatenated across codebooks; otherwise the implementation falls back to codebook-wise token statistics. The descriptor is normalized and used for nearest-neighbor retrieval. These are descriptors in the broad matching sense, but they are learned codec descriptors rather than hand-designed timbre descriptors such as MFCCs or spectral centroid.

The RVQ stack is split into coarse, middle, and fine groups. For a codec with $Q$ codebooks, the boundaries are
\begin{equation}
q_0=\max(1,\lfloor Q/3\rfloor),\quad
q_1=\min(Q,\max(q_0+1,\lfloor 2Q/3\rfloor)).
\end{equation}
For DAC, this yields codebooks 1--3, 4--6, and 7--9. A control $\rho\in[0,1]$ weights the groups by
\begin{equation}
\tilde{w}_c=1-\rho,\quad \tilde{w}_f=\rho,\quad
\tilde{w}_m=(\tilde{w}_c+\tilde{w}_f)/2,
\end{equation}
followed by normalization. The per-grain emission cost is the weighted sum of cosine distances in the three descriptor groups. At runtime, only the top $K$ palette candidates are retained; the DAC evaluation uses $K=96$ and $\rho=0.30$.

\section{Morphing Components}

\subsection{RVQ-Group Transfer}

Two replacement policies are implemented. A full-layer policy swaps all codebooks from the selected palette grain unless a threshold rejects the candidate. The RVQ-group policy treats the stack asymmetrically: the coarse group is copied only when the candidate passes a conservative gate, the middle group is copied from the selected candidate, and the fine group is chosen by a top-$k$ vote over nearby candidates with softmax weights derived from fine-group distances. In the DAC setting, $\theta=0.55$, top-$k=7$, and temperature $\tau=0.47$.

This policy makes the transformation controllable rather than simply more destructive. Coarse tokens tend to carry strong source envelope and event identity information; residual tokens contribute local detail and decoder-facing color. On the tested drum material, the default coarse gate is inactive, so the audible default is best described as gated middle/fine transfer with a coarse safety valve. A forced-band diagnostic confirmed that coarse replacement is acoustically consequential, but it also degrades source envelope preservation, which is why the default avoids unconditional coarse substitution.

For interaction, the group policy exposes a musically meaningful structure/detail axis. Low $\rho$ values bias retrieval toward grains whose coarse and middle descriptors resemble the source, producing conservative palette coloration. Higher $\rho$ values increase the influence of residual detail and can produce more pronounced texture transfer. The gate and dry/wet mix then provide two independent safeguards: the gate protects token regions where no plausible coarse match is found, while dry/wet mixing lets the user reintroduce the original waveform after decoding.

\subsection{Continuity-Constrained Selection}

Independent nearest-neighbor matching often alternates among unrelated palette files, producing audible chattering. We therefore optimize a path through candidate grains. For source grain $n$ and candidate $i$, the objective combines the emission cost $e(i\mid n)$ with a transition cost from the previous selected candidate:
\begin{equation}
J(i_n)=e(i_n\mid n)+\lambda C(i_{n-1},i_n).
\end{equation}
The transition cost rewards adjacent grains from the same file and penalizes file switches and large jumps. The implementation supports greedy matching, greedy smoothing, bounded beam search, and exact Viterbi over the retained candidates. Beam search is the default runtime compromise: it is slower than greedy smoothing, but much cheaper than Viterbi and smoother than frame-local matching.

The distinction matters perceptually. A low emission cost alone can still alternate between files on consecutive grains, especially when a heterogeneous palette contains many short percussive sounds with similar onset descriptors. The transition term encourages locally coherent palette trajectories without requiring the user to curate a perfectly homogeneous corpus. It also makes the result easier to automate: changing beam width or transition strength changes the temporal stability of the palette path while leaving the basic source-to-palette matching objective intact.

Overlapping output grains are merged in token space by deterministic overwrite in increasing grain-start order; the waveform is reconstructed only once by the codec decoder. Thus, the system borrows the selection-and-continuity problem from mosaicing, but the units are learned codec-token grains rather than waveform fragments.

\section{Plugin and Realtime Operation}

\begin{figure}[t]
    \centering
    \includegraphics[width=0.82\columnwidth]{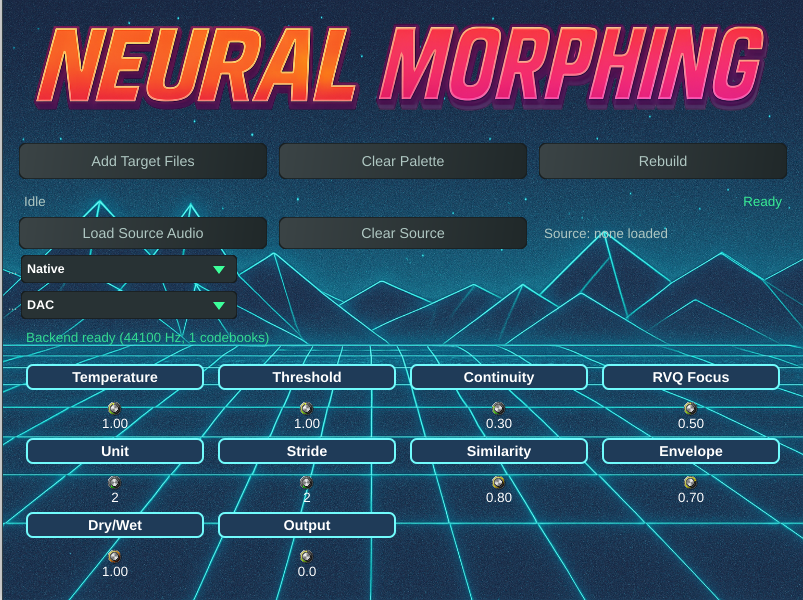}
    \caption{Standalone Neural Morphing interface. The plugin exposes palette selection, dry/wet balance, grain/search parameters, and RVQ transfer controls while hiding codec-token operations.}
    \label{fig:standalone-ui}
\end{figure}

The system is implemented as a JUCE-based standalone/VST3/AU effect plus a Python reference path for ablations and metric extraction. Palette preparation is offline: clips are encoded, segmented, described, and cached. Runtime processing then performs source encoding, descriptor extraction, top-$K$ candidate retrieval, sequence search, RVQ-group swapping, and decoding. Figure~\ref{fig:standalone-ui} shows the current standalone interface used for development and validation.

Realtime operation is bounded rather than sample-synchronous. The plugin stores palette descriptors in memory, optionally uses approximate nearest-neighbor indexing, batches codec calls where possible, and keeps beam width, candidate count, grain size, and RVQ transfer parameters user-adjustable. In live mode, reported latency is
\begin{equation}
\mathrm{latency}_{\mathrm{live}}=\max(256,2B_{\mathrm{blk}})\ \mathrm{samples},
\end{equation}
where $B_{\mathrm{blk}}$ is the host block size. The audio thread does not block on a failed morph update: if the palette is rebuilding, the backend is unavailable, an encode/match/decode stage returns empty, the decoded FIFO overflows, or the wet buffer underruns, the current update is skipped or the oldest wet material is dropped.

Several details are important for using the effect in practice. First, palette loading is deliberately separated from audio processing, because codec encoding and descriptor construction are the most expensive steps when a new corpus is selected. Second, decoded audio is staged through a wet FIFO rather than generated directly inside every host callback. Third, the plugin exposes a dry/wet path so that brief missed wet updates degrade gracefully instead of muting the source. Fourth, the sequence state used by chunked rendering is intentionally small: the previous selected palette index is carried forward, but the full beam lattice is not retained across arbitrary host blocks. These choices favor predictable scheduling over exact equivalence to the offline reference render.

The reference evaluation renders complete excerpts, while the realtime proxy processes sequential chunks and carries the previous palette index into the next chunk. This does not preserve full beam state, so it is not claimed to be identical to full-context rendering. It is a deployment check: it measures the mismatch introduced by chunked operation under plugin-like scheduling.

From the host perspective, the controls are divided into three groups. Palette controls select or rebuild the corpus and determine whether cached descriptors can be reused. Search controls set the candidate count, beam width, grain length, and hop, thereby trading palette-path quality against computation. Transfer controls set dry/wet balance, $\rho$, thresholding, and the RVQ replacement mode. This separation is useful in practice because it maps implementation costs onto user expectations: palette changes may require a rebuild, search changes affect responsiveness, and transfer changes alter the sound with comparatively small scheduling impact once candidates are available.

The implementation also separates monitoring from audio rendering. Backend health flags record whether encoding, candidate retrieval, decoding, channel layout, and determinism checks succeed for a given run. These flags are not presented as musical metrics; they are engineering guards that prevent silent failures from being mistaken for algorithmic behavior. In a production plugin, the same information can be surfaced as status indicators, while the audio path continues with dry signal, cached wet signal, or skipped morph updates when a backend stage is unavailable.

\section{Evaluation}

Evaluation uses deterministic manifests. The strict palette contains 247 Freesound clips of 5--15~s duration with permissive Creative Commons licenses~\cite{fonseca2017freesound}. Source/reference material consists of disjoint 8~s crops from the WaivOps Lo-Fi Drums corpus~\cite{patchbanks_waivops_2025_lofi_drums}. The source/reference clips are comparison material for corpus-proximity metrics, not ground-truth morph targets. Metrics are used only as sanity checks for the demo claims: sequence optimization should reduce unstable palette switching, RVQ grouping should expose a structure/detail trade-off, and chunked operation should remain close enough to full-context rendering to be musically useful.

The evaluation is intentionally aligned with the effect's use case. Percussive material is a demanding setting for chattering because short attacks make discontinuous palette choices audible, and it is also a natural target for rhythm-preserving sound design. We therefore report source-envelope correlation, spectral convergence, log-spectral distance, raw palette-index jitter, path-continuity measures, and realtime factor (RTF). Frechet Audio Distance is included as a set-level descriptive statistic, but it is not treated as a per-example preference metric. All objective measures are interpreted relative to codec reconstruction and held-out corpus proximity rather than exact target recovery.

\begin{table}[t]
    \centering
    \scriptsize
    \setlength{\tabcolsep}{1.7pt}
    \caption{Main DAC ablation on the deterministic source/reference set. Values are means except FAD, which is set-level. Lower is better except envelope correlation. Jit is shown in thousands of raw palette-index jump units.}
    \label{tab:dac-main}
    \begin{tabular}{lrrrrrr}
    \toprule
    Method & FAD & SC & LSD & Jit & EnvC & RTF \\
    \midrule
    Beam RVQ & 1.134 & 1.307 & 27.04 & 11.52k & 0.986 & 0.217 \\
    Beam full & 0.172 & 1.397 & 27.09 & 11.52k & 0.999 & 0.236 \\
    Greedy full & 0.172 & 1.397 & 27.09 & 24.66k & 0.999 & 0.223 \\
    Greedy RVQ & 0.961 & 1.310 & 26.93 & 24.66k & 0.987 & 0.232 \\
    \bottomrule
    \end{tabular}
\end{table}

Table~\ref{tab:dac-main} shows the primary DAC ablation. Beam search halves raw palette-index jitter relative to greedy matching in both swap modes, while RVQ-group transfer changes spectral and envelope diagnostics relative to full-layer replacement. Runtime remains below realtime on the reference path for all four conditions. The numbers should not be read as perceptual preference scores; they confirm that the exposed controls change the intended measurable properties. The full-layer rows are useful as a conservative comparison because, under the default threshold on this slice, they preserve very high envelope correlation. The RVQ rows are the more relevant creative mode because they introduce middle/fine palette detail while retaining source timing.

\begin{table}[t]
    \centering
    \scriptsize
    \setlength{\tabcolsep}{2.4pt}
    \caption{Deployment diagnostics. Top: RVQ-group path-continuity comparison on 96 pairs. Bottom: realtime-proxy parity for 12 chunked renders.}
    \label{tab:deploy}
    \begin{tabular}{lrrrr}
    \toprule
    Selector & Jit & File sw. & Adj. & Seq ms \\
    \midrule
    Greedy & 24.66k & 78.1\% & 14.9\% & 2.5 \\
    Smooth & 13.47k & 40.9\% & 50.3\% & 145 \\
    Beam & 11.52k & 35.2\% & 57.3\% & 1737 \\
    Viterbi & 6.46k & 19.4\% & 76.2\% & 12830 \\
    \bottomrule
    \end{tabular}

    \vspace{0.35em}
    \begin{tabular}{rrrr}
    \toprule
    Chunk & SC & LSD & EnvCorr \\
    \midrule
    8192 & 0.355 & 10.60 & 0.983 \\
    16384 & 0.317 & 9.28 & 0.986 \\
    32768 & 0.291 & 8.68 & 0.988 \\
    \bottomrule
    \end{tabular}
\end{table}

The path-continuity rows in Table~\ref{tab:deploy} show why beam search is used in the demo: it reduces file switching and increases adjacent within-file transitions while staying far below Viterbi runtime. The chunked parity rows show that larger chunks move closer to the full-context render on spectral metrics while envelope correlation remains high. A palette-scaling sweep over 32, 64, 128, and 247 clips gave median realtime factors between 0.229 and 0.237 for beam+RVQ-group on the diagnostic split, although total runtime has a heavy tail when palette encoding and cache effects dominate. These results support deployment under bounded palette and backend assumptions, not universal realtime guarantees.

A separate grain/hop diagnostic gave similar envelope correlations across $G/H$ settings, while sequence runtime decreased as hops became larger. This is expected: short hops provide more opportunities for local correction but increase the number of matching decisions, whereas longer hops reduce search load at the cost of coarser temporal control. For the plugin, the $G=7,H=2$ DAC setting is a compromise between responsive edits, stable envelope preservation, and sequence runtime. The same design principle applies to other codec integrations, although frame rate and token layout change the best numerical setting.

The results also clarify which parameters are safe for live interaction. Dry/wet, $\rho$, transfer policy, and threshold are expressive controls because they act after candidate retrieval or reuse computed descriptors. Palette replacement and codec switching are heavier and are therefore treated as state changes rather than continuous performance gestures. Beam width and candidate count sit between these extremes: they alter path stability but directly affect search cost. The interface therefore favors a few interpretable presets rather than exposing every internal constant.

\section{Discussion and Use}

Neural Morphing is intended as a sound-design instrument rather than an offline target-reconstruction algorithm. In a typical scenario, the user loads a drum or percussive source, chooses a palette of environmental, instrumental, or synthetic sounds, and adjusts dry/wet, grain size, beam width, and RVQ structure/detail controls until the source rhythm acquires palette color. The system is most useful when the palette is treated as material rather than as a fixed target: found sounds can add noisy residual detail to a clean drum track, while synthesized transients can impose a sharper or more electronic character. Because the method is training-free, these changes can be made by swapping the palette rather than collecting a dataset and retraining a model. The strongest current evidence is for percussive material, where onset preservation and palette chattering are easy to inspect. Harmonic, vocal, and long-form material may require different gates, descriptors, or transition models.

The demonstration workflow follows the same logic. First, the dry source establishes the rhythmic reference. Second, greedy and beam selection are compared with the same palette so that the effect of path continuity can be heard directly. Third, full-layer and RVQ-group transfer are compared to show the difference between broad replacement and controlled residual coloration. Finally, chunk size or search width exposes the responsiveness/quality trade-off encountered in a DAW. This makes the components audible without requiring inspection of token streams or objective metrics.

The main limitation is perceptual validation. Objective diagnostics can show continuity, runtime, codec health, and source/palette proximity, but they cannot establish listener preference or musical usefulness. Direct auditioning is therefore central: listeners can compare greedy and sequence-aware selection, full-layer and RVQ-group transfer, and the responsiveness/chunk-size trade-off. Future listening tests should separate preference, perceived continuity, source-rhythm preservation, palette recognizability, and tolerance of neural-codec artifacts.

\section{Conclusion}

This paper presented Neural Morphing, a deployable codec-token audio effect for palette-based morphing. The retained components are an RVQ-group transfer policy, a continuity-constrained candidate selector, and a bounded VST3/AU implementation that supports chunked operation. DAC experiments show that beam search reduces palette-path instability, RVQ grouping exposes a useful structure/detail axis, and the reference path remains below realtime under the tested palette sizes. Future work will expand beyond percussive material, add controlled listening tests, and improve lower-latency streaming while preserving the training-free workflow.

\section{Acknowledgements}
This work is supported by the European Research Council (ERC) under the EU's Horizon 2020 research \& innovation programme, grant agreement No.\ 101019375 (\textit{Whither Music?}).

\bibliographystyle{IEEEtranDAFx}
\bibliography{DAFx26_tmpl}

\end{document}